%% file: conference_101719.tex
\documentclass[conference]{IEEEtran}
\IEEEoverridecommandlockouts
\usepackage{cite}
\usepackage[numbers]{natbib}
\usepackage{amsmath,graphicx,booktabs,arydshln}
\usepackage{amsmath,amssymb,amsfonts}
\usepackage{algorithmic}
\usepackage{graphicx}
\usepackage{textcomp}
\usepackage{color}
\usepackage[table,x11names]{xcolor}
\definecolor{Gray}{gray}{0.9}
\newcommand\addgitlink{\texttt{https://kinwaicheuk.github.io/IJCNN2021.github.io/}}
\def\BibTeX{{\rm B\kern-.05em{\sc i\kern-.025em b}\kern-.08em
    T\kern-.1667em\lower.7ex\hbox{E}\kern-.125emX}}

\begin{document}

\title{Revisiting the Onsets and Frames Model with Additive Attention\\
}

\author{\IEEEauthorblockN{Kin Wai Cheuk}
\IEEEauthorblockA{\textit{Information Systems,}\\
	\textit{Technology, and Design}\\
	\textit{Singapore University}\\
	\textit{of Technology and Design}\\\\
	\textit{Institute of High Performance}\\
	\textit{Computing, A*STAR}\\
kinwai\_cheuk@mymail.sutd.edu.sg\\
}
\and
\IEEEauthorblockN{Yin-Jyun Luo, Emmanouil Benetos}
\IEEEauthorblockA{\textit{School of Electronic Engineering }\\
    \textit{and Computer Science,}\\
	\textit{Queen Mary University of London}\\
yin-jyun.luo@qmul.ac.uk\\
emmanouil.benetos@qmul.ac.uk}
\and
\IEEEauthorblockN{Dorien Herremans}
\IEEEauthorblockA{\textit{Information Systems,}\\
	\textit{Technology, and Design}\\
	\textit{Singapore University}\\
	\textit{of Technology and Design}\\\\
dorien\_herremans@sutd.edu.sg}
}

\maketitle

\begin{abstract}
Recent advances in automatic music transcription (AMT) have achieved highly accurate polyphonic piano transcription results by incorporating onset and offset detection. The existing literature, however, focuses mainly on the leverage of deep and complex models to achieve state-of-the-art (SOTA) accuracy, without understanding model behaviour.
In this paper, we conduct a comprehensive examination of the Onsets-and-Frames AMT model, and pinpoint the essential components contributing to a strong AMT performance. 
This is achieved through exploitation of a modified additive attention mechanism.
The experimental results suggest that the attention mechanism beyond a moderate temporal context does not benefit the model, and that rule-based post-processing is largely responsible for the SOTA performance.
We also demonstrate that the onsets are the most significant attentive feature regardless of model complexity.
The findings encourage AMT research to weigh more on both a robust onset detector and an effective post-processor.

\end{abstract}


\section{Introduction}
\label{sec:intro}

Automatic music transcription (AMT) has been a crucial task in music information retrieval (MIR) that underlies a variety of important applications, such as turning a mass of audio data to an indexable format which enables queries based on musical structure~\cite{cuthbert2010music21}, converting the audio to a symbolic dataset taken as the input for music generation~\cite{huang2020pop, herremans2017functional} or music accompaniments to play along with~\cite{magalhaeschordify}.

Existing literature has been focusing on extending network capacity to achieve state-of-the-art (SOTA) transcription accuracy.
This includes fully convolutional neural networks~\cite{springenberg2014striving}, hybrid convolutional and recurrent neural networks~\cite{Sigtia2015AnEN}, and convolutional sequence-to-sequence models~\cite{Sigtia2015AnEN}.
In parallel to increasing model complexity, incorporating onset~\cite{Hawthorne2017OnsetsAF} and offset~\cite{kim2019adversarial, kelz2019deep} detection, and leveraging a large dataset~\cite{hawthorne2018enabling} for model training are also shown to improve the performance.
Despite the development, the components responsible for the superior performance remain unclear. 
To the best of our knowledge, only Kelz \textit{et al.} have attempted to explain AMT models using invertible neural networks~\cite{Kelz18}.
Although the work hints towards how the model possibly captures the notion of musical notes, it does not provide further insights on the relevant features for transcription.

The main proposition of this paper is to elucidate the underlying components that lead to a performant AMT model that contribute to achieving SOTA transcription accuracy.
We aim to anlayze and identify 1) the feature on which Onsets and Frames~\cite{Hawthorne2017OnsetsAF} relies the most to achieve the SOTA accuracy; 2) the length of temporal context with which the classifier obtains the most gain in accuracy; and 3) the interplay between the temporal information and different model constitutions.
These are achieved by using the additive attention~\cite{Bahdanau2015NeuralMT} which is slightly modified to facilitate our analysis.
The results indicate that, although temporal information is helpful, the length of the attentive context has to be limited in order to obtain a superior performance. 
Additionally, given a decent accuracy of onset prediction, the rule-based inference model accounts for the majority of the improvement in terms of note-wise transcription accuracy.
Our findings shed lights on promising avenues of research for improving AMT systems.

We structure the rest of the paper as follows.
The background relevant to Onsets and Frames is provided in Section~\ref{sec:related_works}.
In Section~\ref{sec: proposed_work}, we propose the framework for analyzing Onsets and Frames, using the modified additive attention~\cite{Bahdanau2015NeuralMT} which serves as the probing tool.
Section~\ref{sec: experiment} elaborates the experimental setups including the dataset, model parameters, and the evaluation methods.
We thoroughly discuss the experimental results in Section~\ref{sec:result}, and conclude the paper in Section~\ref{sec:conclusion}.

\begin{figure*}[htb]
    \centering
    \centerline{\includegraphics[width=1.0\linewidth]{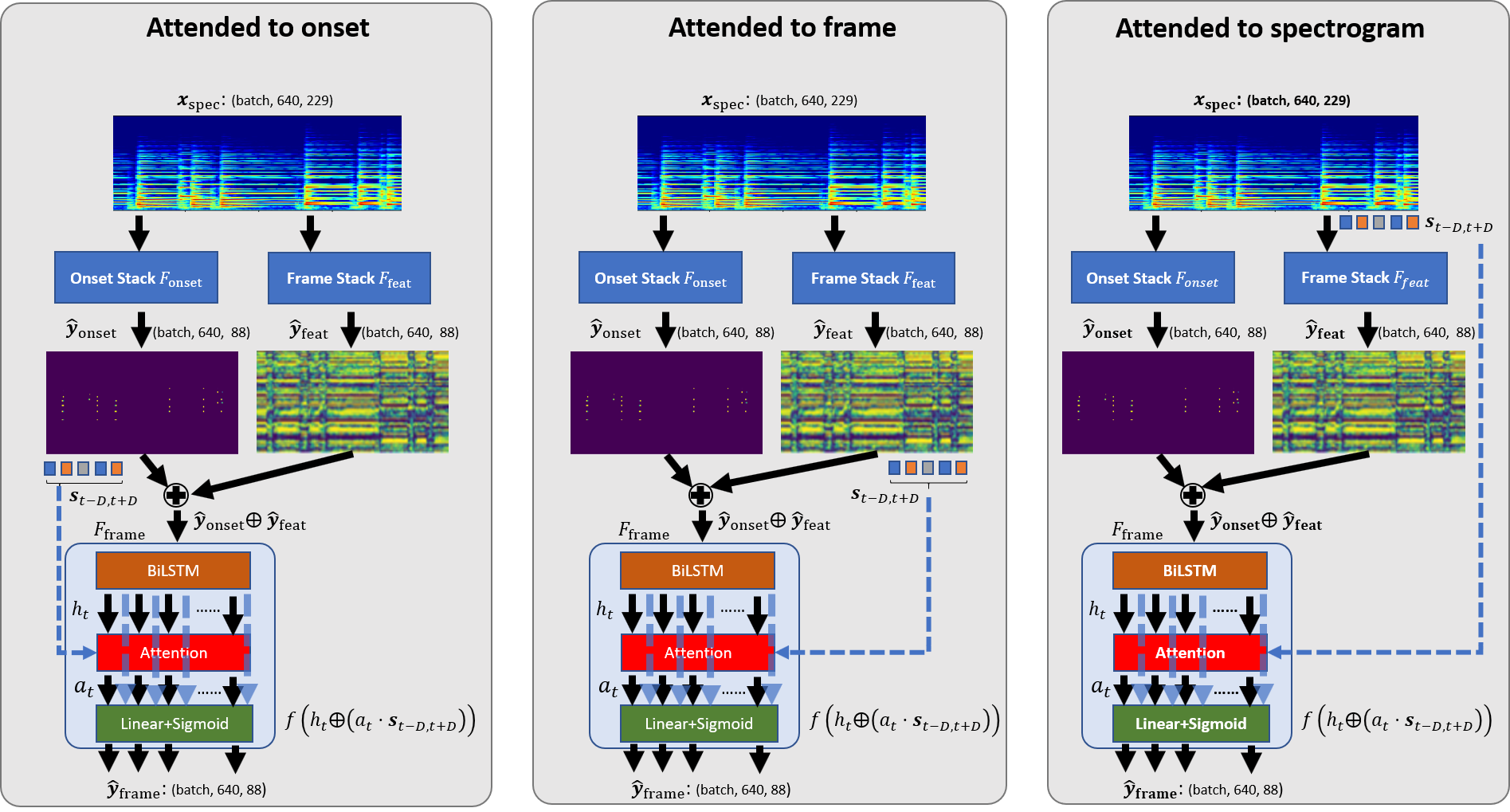}}
    \caption{The schematic diagram showing the use of additive attention mechanism to study the Onsets and Frames model. Different parts of the model are being attended, and the results are reported in Table~\ref{tab:compare}.}\medskip
    \label{fig:proposed_model}
\end{figure*}
\section{Background}
\label{sec:related_works}
Onsets and Frames is a model which performs both onset location prediction and frame-wise multi-pitch detection~\cite{Hawthorne2017OnsetsAF}. These two outputs are then used during inference to achieve state-of-the-art piano transcription accuracy. This model contains three major stacks, namely, an onset prediction stack $F_{\text{onset}}$ (consisting of four convolutional blocks, one bi-directional long short-term memory (biLSTM) layer, and one fully connected layer), a feature extraction stack $F_{\text{feat}}$ (four convolutional blocks and one fully connected layer), and a frame prediction stack $F_{\text{frame}}$ (four convolutional blocks, one biLSTM layer, and one fully connected layer) as follows:
\begin{equation}
\label{eq:onset_frame}
\left\{
\begin{aligned}
    \boldsymbol{\hat{y}_{\text{onset}}}&= F_\text{onset}({\boldsymbol{x_{\text{spec}}}})\\ 
    \boldsymbol{\hat{y}_{\text{feat}}}&= F_{\text{feat}}(\boldsymbol{x_{\text{spec}}})\\ 
    \boldsymbol{\hat{y}_{\text{frame}}}&= F_{\text{frame}}(\boldsymbol{\hat{y}_{\text{onset}}}\oplus \boldsymbol{\hat{y}_\text{feat}})
\end{aligned}
\right.
\end{equation}
where $\boldsymbol{x_{\text{spec}}} \in [0,1]^{T\times N}$ is the normalized log-magnitude spectrogram with number of timesteps $T$ and number of bins $N$; $\boldsymbol{\hat{y}_{\text{onset}}}\text{, } \boldsymbol{\hat{y}_{\text{frame}}} \in [0,1]^{T\times 88}$, and $\boldsymbol{\hat{y}_{\text{feat}}} \in \mathbb{R}^{T\times 88}$ are the outputs from different stacks $F$. The concatenated outputs $\boldsymbol{\hat{y}_{\text{onset}}}\oplus \boldsymbol{\hat{y}_\text{feat}}$ are used as the input to the $F_{\text{frame}}$ stack. The objective function $L$ to be minimized during training consists of two binary cross-entropy loss components for onsets and frames as:
\begin{equation}
\label{eq:objective}
    L = \text{BCE}(\boldsymbol{\hat{y}}_{\text{onset}},\boldsymbol{{y}}_{\text{onset}}) + \text{BCE}(\boldsymbol{\hat{y}}_{\text{frame}}, \boldsymbol{{y}}_{\text{frame}})
\end{equation}
where $\boldsymbol{y}_{\text{onset}}$ and $\boldsymbol{y}_{\text{frame}}$ are the onset and multi-pitch activation ground-truth labels. The final pianoroll prediction $\hat{\boldsymbol{y}}_{\text{roll}}\in \{0,1\}^{T\times 88}$ is obtained via a rule-based interface function $g$:
\begin{equation}
    \hat{\boldsymbol{y}}_{\text{roll}} = g(\hat{\boldsymbol{y}}_{\text{onset}}, \hat{\boldsymbol{y}}_{\text{frame}})
\end{equation}
that outputs a ``note on'' event only when the frame activation comes with an onset. The transcription accuracy is calculated from the $\hat{\boldsymbol{y}}_{\text{roll}}$, $\boldsymbol{{y}}_{\text{frame}}$ pair instead of the $\hat{\boldsymbol{y}}_{\text{frame}}$, $\boldsymbol{{y}}_{\text{frame}}$ pair.

\section{Methodology}
We describe our proposed methodology for answering the research questions in this section, along with the modified additive attention mechanism used for the study.

\label{sec: proposed_work}

\subsection{Research Questions}\label{sec:rq}
As mentioned in Section~\ref{sec:intro}, we aim to answer 
1) which feature ($\boldsymbol{x_{\text{spec}}}$, $\boldsymbol{\hat{y}}_{\text{feat}}$, or $\boldsymbol{\hat{y}}_{\text{frame}}$) does the final classifier rely on most in the Onsets-and-Frames model; 
2) how much temporal information is required for the classifier to achieve a high transcription F1-score; and 
3) how does the temporal information, induced by the attention mechanism, interact with different network components, and affect the transcription performance.
Our analysis based on the additive attention mechanism proposed by Bahdanau \textit{et al.}~\cite{Bahdanau2015NeuralMT}.
Attention is considered as an add-on to LSTMs, which provides model interpretability through attentive feature maps.
We choose this particular attention mechanism to minimize the modification to the original Onsets-and-Frames model.

More specifically, to answer \textit{question 1}, we add the attention module to which different input features are presented, and evaluate the corresponding accuracy of transcription.
The attentive feature that corresponds to the best performance is potentially the important feature on which Onsets and Frames relies on.
Visualization of the attentive feature maps also sheds light on the most significant feature responsible for the transcription.
The experimental results are detailed in Section~\ref{subsec: exp1}.

In order to answer \textit{question 2}, we constrain model capacity and simply use a linear layer coupled with attention.
The constraint is to assure that the temporal information is only accessible by the model through the attention mechanism.
This facilitates our analysis because the performance difference under this setup is only attributable to the context length of the attentive features, avoiding confounding factors, whereby we can more explicitly evaluate the effect of length of temporal information on the transcription accuracy.
Figure~\ref{fig:win_size} from Section~\ref{subsec: D_size} shows the corresponding results.

For \textit{question 3}, we conduct a comprehensive ablation study to thoroughly examine interactions between the attention mechanism and individual model components in Onsets and Frames.
Specifically, we remove bit by bit the onset stack, the biLSTM layers, the convolutional layers, the attention mechanism, and the inference model, and observe the corresponding change in transcription accuracy.
This helps elucidate the interplay between each individual component, and the extent to which the temporal information improves performance.
The results are reported in Section~\ref{subsec: components}. 
We note that Hawthorne \textit{et al.}~\cite{Hawthorne2017OnsetsAF} also conducted a similar ablation study, and we will highlight the differences and distinguish ourselves from their study in Section~\ref{subsec: components}.

\subsection{Modified Additive Attention}
We adapt the additive attention~\cite{Bahdanau2015NeuralMT} to our analysis and describe the modification as follows.
The original attention mechanism posits a challenge to our limited computational resource. 
In particular, each input sequence of our dataset corresponds to 640 timesteps under the experimental configuration, making it prohibitively expensive to use the global attention proposed by Bahdanau \textit{et al.}~\cite{Bahdanau2015NeuralMT} which was designed for dozens of timesteps.
We thereby modify and obtain the \emph{local attention mechanism} similar to Luong \textit{et al.}~\cite{luong2015-effective} as follows:
\begin{equation}
\label{eq:attention1}
\boldsymbol{a_t} = \texttt{softmax}\bigl(\boldsymbol{v}\ \texttt{tanh(\texttt{attn}}(h_{t}, {\boldsymbol{s}}_{t-D, t+D}))\bigl)
\end{equation}
where \texttt{attn} denotes the attention mechanism~\cite{Bahdanau2015NeuralMT}, $\boldsymbol{v}$ is the weight for the linear layer reducing the feature dimension to $1$, and $\boldsymbol{a_t} \in [0,1]^{(2D+1) \times 1}$ is the attention score with a local window size of $2D+1$. 
The input ${\boldsymbol{s}}_{t-D, t+D} \in \mathbb{R}^{(2D+1)\times N_{\text{feat}}}$ is the sequence (either $\boldsymbol{x_{\text{spec}}}$, $\boldsymbol{\hat{y}}_{\text{feat}}$, or $\boldsymbol{\hat{y}}_{\text{frame}}$) to which the attention applied, covering $D$ timesteps before and after the current timestep $t$. 
$h_{t}$ is the hidden state of a biLSTM network prior to the final classification layer $f$ (the green block in Figure~\ref{fig:proposed_model}) allocated in the frame stack $F_{\text{frame}}$. 

In order to pinpoint the significant features responsible for the final predictions, we couple the attention with the the classification layer $f$ as:
\begin{equation}
\label{eq:myattention}
    \boldsymbol{\hat{y}}^{t}_{\text{frame}}= f(h_{t}\oplus (\boldsymbol{a_t} \cdot \boldsymbol{s}_{t-D, t+D})).
\end{equation}

\begin{figure*}[htb]
    \centering
    \centerline{\includegraphics[width=1.0\linewidth]{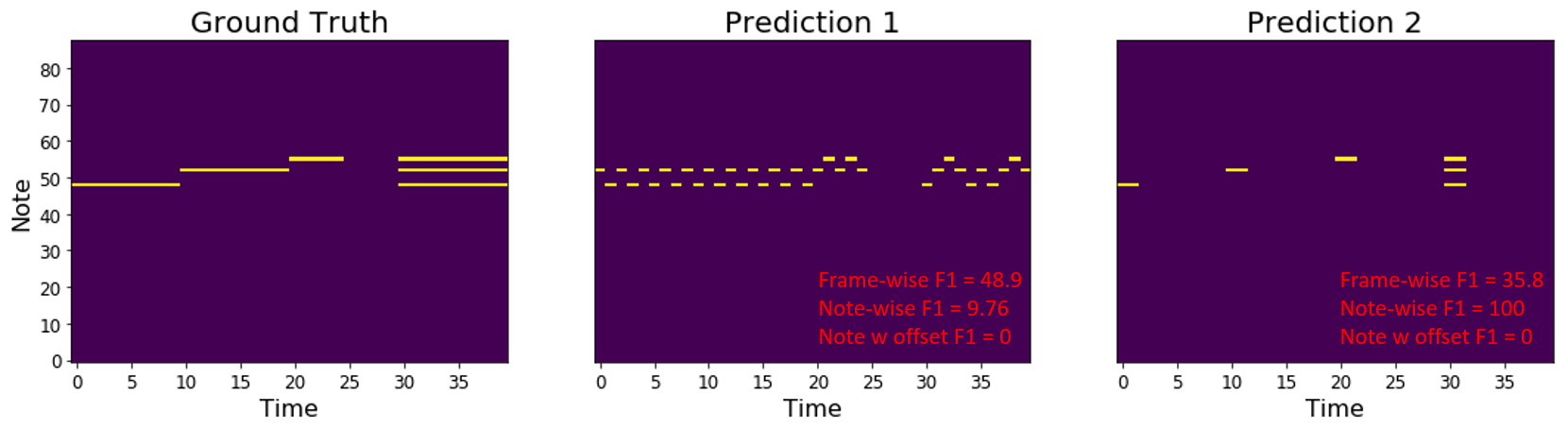}}
    \caption{Differences between frame-wise, note-wise, and note-with-offset-wise metrics. \textit{Prediction 2} is closer to the ground truth in terms of musical structure, yet, it has a lower frame-wise F1 then \textit{Prediction 1}. Therefore reporting frame-wise F1 alone is misleading.}\medskip
    \label{fig:metrics}
\end{figure*}

The schematic diagram of our models is shown in Figure~\ref{fig:proposed_model}. Following Bahdanau \textit{et al.}~\cite{Bahdanau2015NeuralMT}, attention is applied to the time-axis, as we focus on analyzing the temporal dimension in this work.
Future research could also investigate more advanced attention mechanisms which consider both the time- and frequency-axes such as the one proposed by Xu \textit{et al.}~\cite{XuBKCCSZB15}. 

\section{Experiments}
\label{sec: experiment}

\subsection{Dataset}
\label{subsec: dataset}
We train and evaluate our model with the MAPS dataset~\cite{emiya2010maps}. We follow the same training and test splits as in the existing literature~\cite{Sigtia2015AnEN,Hawthorne2017OnsetsAF} by removing pieces in the training set that are also present in the test set, leaving only 139 training recordings and 60 test recordings. All audio recordings from the datasets are downsampled to $16$~kHz, and then Mel spectrograms are extracted from these recordings using a Hann window size of $2048$, hop size of $512$, and $229$ Mel bins. It has been shown in the literature that the Mel spectrogram outperforms other spectral representations in the context of deep learning-based AMT~\cite{Hawthorne2017OnsetsAF, cheuk2020impact, Kelz2016OnTP}.

\subsection{Implementation}
The work is based on \texttt{PyTorch}, and we use the adapted implementation of Onsets and Frames\footnote{\texttt{https://github.com/jongwook/onsets-and-frames}}, originally implemented in \texttt{TensorFlow} for our experiments.
We use the same Adam optimizer as in Hawthorne \textit{et al.}~\cite{Hawthorne2017OnsetsAF} but slightly change the learning rate to $6\times 10^{-5}$ since it shows a faster model convergence in our experiments. 
To ensure convergence, we train our model for 20,000 epochs which is equivalent to 160,000 steps with a batch size of 16. All spectrograms are extracted on-the-fly with nnAudio~\cite{cheuk2019nnaudio}. 


\subsection{Evaluation Metrics}
\subsubsection{Frame-wise metric} 
Frame-wise accuracy, despite being commonly adopted in the literature, is a naive metric which calculates accuracy by comparing a prediction with the ground-truth pianoroll in a pixel-by-pixel fashion.
This metric has shown to not correlate well with perceptual quality~\cite{Hawthorne2017OnsetsAF, bay2009evaluation}.

Figure~\ref{fig:metrics} shows an example where a high frame-wise score could have an inferior perceptual quality of transcription.
The ground-truth pianoroll on the left shows three successive notes \texttt{C}, \texttt{E}, and \texttt{G} within the interval from 0- to 25-th timestep.
The interval of 30- to 35-th timestep highlights a \texttt{C major} chord which consists of another three notes (\texttt{C}, \texttt{E}, \texttt{G}) which amounts to six notes in total.
\textit{Prediction 1} at the middle obtains a higher frame-wise F1-score than \textit{Prediction 2} on the right, due to the fact that the former captures the pixels more accurately.
\textit{Prediction 2}, however, is more perceptually relevant, attributed to the correct prediction of the number of notes.

\subsubsection{Note-wise metric}
Following the discussion above, one can expect that note-wise metrics correlate better with perceptual quality, which evaluates the prediction on a note-by-note basis~\cite{Hawthorne2017OnsetsAF}.

With note-wise metrics, a correct prediction should be at the ground-truth pitch and onset with a tolerance of 50ms.
As mentioned earlier, \textit{Prediction 2} yields a perfect note-wise F1-score as it matches exactly to the ground-truth pianoroll in terms of the total number of notes, and meets the criteria at the same time.
On the other hand, \textit{Prediction 1} matches by only two notes, \texttt{E} and \texttt{C} at 20- and 30-th timestep, respectively, resulting in a recall as low as $33$.
Due to a large amount of wrong predictions, the precision drops to $5.71$, resulting in a low F1-score of $9.76$.
Therefore, the note-wise metric is more musically sensible than the frame-wise metric.

\subsubsection{Note-with-offset-wise metric}
The note-with-offset-wise metric extends the note-wise metric by also considering the note offset, with a tolerance of $50$ms or $20\%$ of the note duration, whichever is larger~\cite{bay2009evaluation}.
This metric thereby takes into account the transcribed note duration additionally.
Since the predicted lengths of the notes in both \textit{Prediction 1} and \textit{Prediction 2} deviate much from the ground-truth annotations, the F1-score for this metric is 0 for both cases.

We use the implementations from \texttt{mir\_eval}\footnote{\texttt{https://github.com/craffel/mir\_eval}} to calculate and report the above-mentioned metrics; 
specifically, \texttt{mir\_eval.multipitch.evaluate} for frame-wise, \texttt{mir\_eval.transcription.evaluate\_notes} for both frame-wise and note-with-offset-wise metric (differentiated with the argument \texttt{offset\_ratio}).

\section{Results}
\label{sec:result}
\subsection{Onsets and Frames with Attention Mechanism}
\label{subsec: exp1}
\input{Tables/compare}

As mentioned in Section~\ref{sec:rq}, we couple Onsets and Frames with attention, whereby we analyze the responsible feature for the performance.
Table~\ref{tab:compare} shows the transcription results for the models with and without attention (baseline). 
The Wilcoxon signed-rank test on the recording-level F1-scores shows that when attending to $\hat{\boldsymbol{x}}_{\text{spec}}$, our model attains significant improvement over the baseline in terms of both frame-wise ($p = 0$) and note-wise metrics ($p=0.018$).
Attending to $\hat{\boldsymbol{x}}_{\text{onset}}$
yields significant improvements in terms of frame-wise ($p=0$) and note-with-offset-wise ($p=0.016$) F1-scores.
On the other hand, applying the attention to $\hat{\boldsymbol{x}}_{\text{feat}}$ only significantly improves the frame-wise metric ($p=0.003$). 
Accordingly, using $\hat{\boldsymbol{x}}_{\text{spec}}$ or $\hat{\boldsymbol{x}}_{\text{onset}}$ as the attentive feature outperforms $\hat{\boldsymbol{x}}_{\text{feat}}$, which hints towards the significant features in Onsets and Frames could indeed be the note onsets.
We will discuss the contributions from other components such as the rule-based post-processor in the later section.

Note that although our aim throughout this paper is not achieving SOTA performance, the augmentation of the attention mechanism does have significant effects according to the statistical test.
One reason for the rather incremental improvements is that the hidden states $h_t$ from the biLSTM might already contain the necessary temporal information for the final classifier $f$, which is supported in rows 7 and 11 of Table~\ref{tab:inference} that the attention can boost the performance in the absence of the biLSTM layer. 
The reason for the attention not being able to serve as a drop-in replacement requires further investigation. 
In addition to biLSTM, the convolutional layers allocated in each stack $F_{\text{onset}}$, $F_{\text{feat}}$, and $F_{\text{frame}}$ can also extract temporal features with the kernel. 
Therefore, the benefit brought from the attention might be overshadowed by both LSTM and convolutional layers.

\begin{figure*}[t]
    \centering
    \centerline{\includegraphics[width=1.0\linewidth]{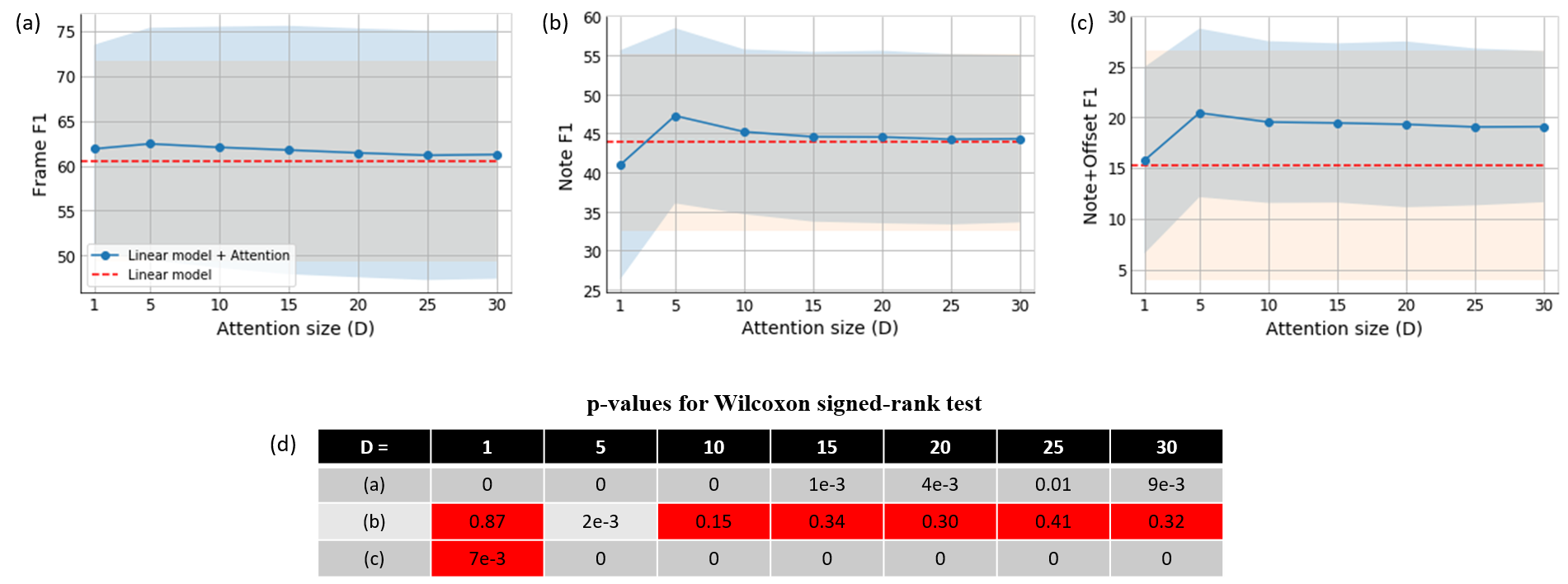}}
    \caption{The F1-scores for frame-wise, note-wise, note-with-offset-wise metrics with different attention sizes $D$. The shaded area represents the standard deviation.}\medskip
    \label{fig:win_size}
\end{figure*}

\subsection{Effect of Attention Size}
\label{subsec: D_size}
The purpose of this experiment is to identify the amount of temporal duration that is necessary for a high transcription F1-score.
As mentioned in Section~\ref{sec: proposed_work} and~\ref{subsec: exp1}, LSTM and convolutional layers could interfere with the attention mechanism, we thus remove them and constrain the model to rely only on temporal information introduced by the attention mechanism. 

Specifically, the model used in this experiment is simplified as a single linear layer $f$ which takes as input the attentive feature: 
\begin{equation}
\label{eq:attention2}
    \boldsymbol{\hat{y}}^{t}_{\text{frame}}= f_D\biggl(\sum_{t=t-D}^{t+D} a_t \cdot \boldsymbol{x}_{\text{spec}(t)}\biggl).
\end{equation}

We experiment with different window sizes of attention $D = \{1, 5, 10, 15, 20, 25, 30\}$, and report the corresponding performances in Figure~\ref{fig:win_size}.
It shows that in the single-layer model, the attention mechanism can significantly improve the performance according to the Wilcoxon signed-rank test. 
Specifically, F1-scores for frame-wise and note-with-offset-wise metrics are improved across most of the cases, while the note-wise metric is only reported significant at $D=5$.
The corresponding $p$-values are reported in panel $d$ of Figure~\ref{fig:win_size}, with red cells indicating the failure of rejecting the null hypothesis. 

\begin{figure*}[htb]
    \centering
    \centerline{\includegraphics[width=1.0\linewidth]{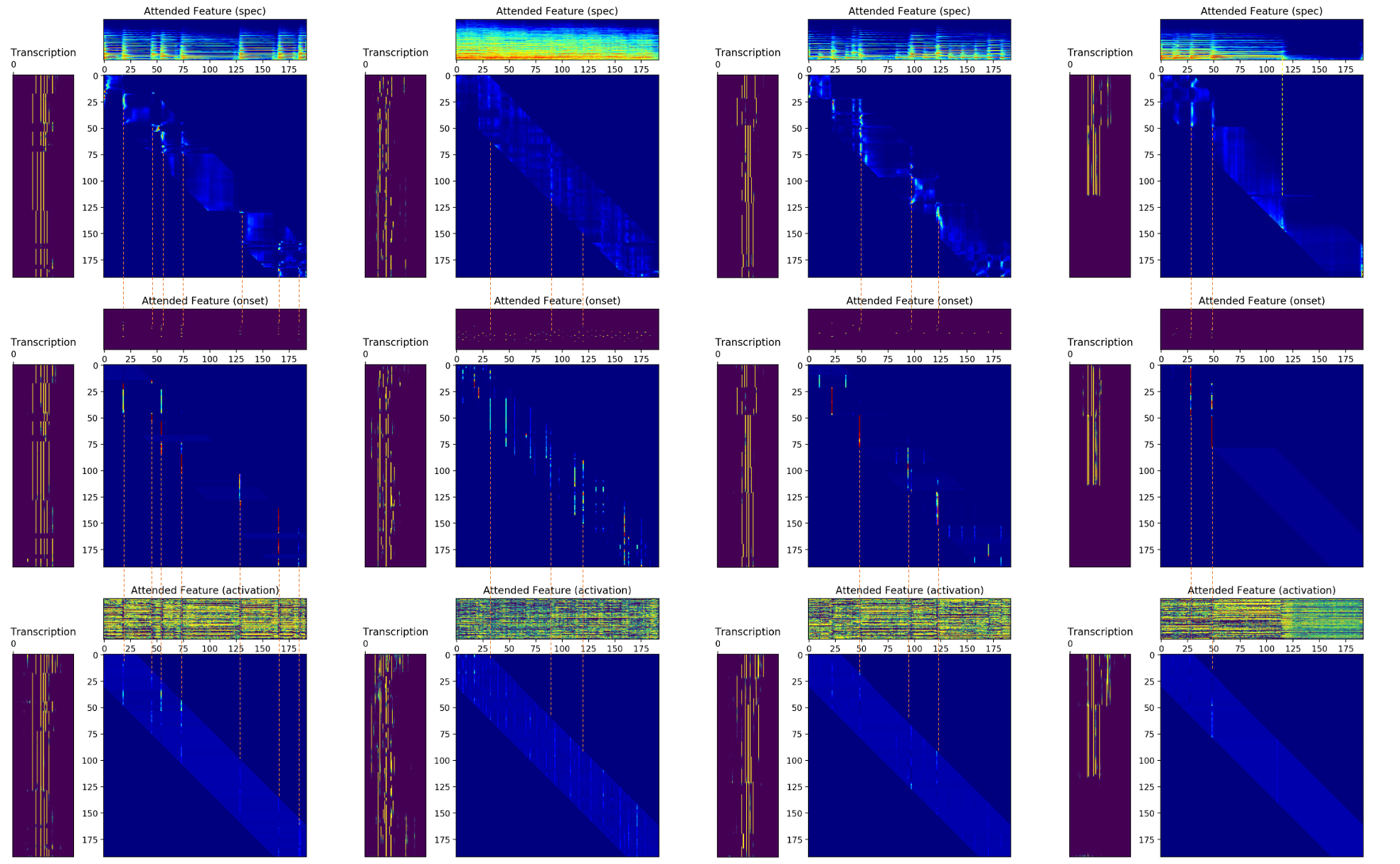}}
    \caption{Attention maps between input sequence ($\hat{\boldsymbol{x}}_{\text{spec}}$, $\hat{\boldsymbol{y}}_{\text{onset}}$, and $\hat{\boldsymbol{y}}_{\text{feat}}$) and the output sequence $\hat{\boldsymbol{y}}_{\text{frame}}$. Best viewed in color.}\medskip
    \label{fig:attention_map}
\end{figure*}

$D = 1$ amounts to having an attentive window of three timesteps, centered at $t$; the short context does not make much difference in terms of note-wise and note-with-offset-wise metrics compared to the baseline model without attention.
It is, however, interesting to find that a longer attention window is not necessarily beneficial, and $D=5$ (around $0.16$ seconds) is shown to be the sweet spot.
This is possibly due to, as we can deduce from the attention map shown in Figure~\ref{fig:attention_map}, that a large-size attention window might confuse the model when excerpts with high note density are presented.
That is, the attention weights are distributed across a relatively large number of notes, turned ``smeared'' along the time-axis.
We will discuss further in Section~\ref{subsec: attention_maps}.

This experiment shows that AMT models require a moderate amount of temporal information, too much or too little might result in a non-optimal performance. 

\input{Tables/inference}

\subsection{Effect of Different Modules}
\label{subsec: components}
As mentioned in Section~\ref{sec:rq}, we aim to study the interplay between each model component in this experiment.
Note that Hawthorne \textit{et al.}~\cite{Hawthorne2017OnsetsAF} also carried out an ablation study for Onsets and Frames. 
We distinguish our study from them by also introducing the attention mechanism, and evaluating models with constrained capacity.
We believe that this approach helps elucidate the interactions between different model constituents.

For a fair benchmark in our analysis, we include our \texttt{PyTorch} re-implementation of the ablation study~\cite{Hawthorne2017OnsetsAF} and the corresponding performance in Table~\ref{tab:inference}.
In particular, rows 13, 14, and 16 are the results for the \texttt{PyTorch} implementation and they correspond to the ablation study (a), (b), and (g) reported in Hawthorne \textit{et al.}~\cite{Hawthorne2017OnsetsAF}, respectively.

\subsubsection{Differences between implementations}
Although our implementation is quite close to the original paper in terms of F1-score for this particular experiment, there are subtle differences.
Rows 13 and 14, for example, are slightly different from the original paper. In the original paper, the note-with-offset-wise F1-scores are seriously impaired when the onset stack or the inference module is removed. While our experiments show less severe impairments in terms of note-with-offset-wise F1-scores. One reason is that we did not use the weighted cross entropy as the original implementation. The weighted cross entropy puts higher weights on onsets and smaller weights on note sustain and offsets; it can help train the model to make a more accurate prediction for the onsets, but potentially cause a relatively stronger deterioration to the offset predictions as the model gets worse by removing some of the stacks. Nonetheless, it does not affect our following discussions.

\subsubsection{biLSTM and attention mechanism (Rows 9-16)} The 15th row is our new experiment where we remove all the biLSTM layers from both the onset stack $F_{\text{onset}}$ and the final frame stack $F_{\text{frame}}$, but do not share the weight for both stacks. We would like to know how strongly would the temporal information affect the model performance. 

When applying the attention mechanism on the full Onsets-and-Frames model (row 12 and row 16 of Table~\ref{tab:inference}), the attention mechanism only improves the F1-scores for frame-wise, note-wise, and note-with-offset-wise by $1.62\%$, $0.51\%$, and $1.23\%$, respectively. When the biLSTM layers are removed (row 11 and row 15 of Table~\ref{tab:inference}), the improvements are $2.03\%$, $1.20\%$, and $2.54\%$, respectively. When we keep biLSTM layers and remove the onset stack $F_{\text{onset}}$ or the onset inference module (rows 9-10 and rows 13-14 of Table~\ref{tab:inference}), the attention mechanism does not improve the transcription accuracy. 

These results indicate that in the absence of biLSTM, the attention mechanism can provide useful temporal information for the model to improve the transcription accuracy. Temporal information can be extracted not only by the biLSTM layers, but also convolutional layers with kernel size greater than $1$ along the time-axis. Therefore, when a model contains deep convolutional layers and biLSTM, the benefit of the attention mechanism is overshadowed. To prove this, we conducted another set of experiments to show that when the attention mechanism is the only source of temporal information, the attention mechanism itself can improve the transcription accuracy greatly.

\subsubsection{Temporal information (Rows 1-8)} Since both recurrent and convolutional neural networks (when the kernel size along the time dimension is greater than $1$) have the ability to extract temporal information, we would like to verify if the attention mechanism would be more beneficial when it is the only source of temporal information. A single layer frame-wise linear model, and a frame-based convolutional model consists of a convolutional layer with kernel size $1\times3$ and a classifier layer outputting a dimension of $88$ are good choices for this experiment. For linear layers (rows 1 and 3), adding the attention mechanism improves frame-wise, note-wise, and note-with-offset-wise F1-scores by $3.14\%$, $7.76\%$, and $34.2\%$, respectively. Similarly, the frame-based convolutional models (rows 5 and 7) gain $4.7\%$, $7.13\%$, and $27.6\%$, respectively when attention is applied. When the model only has access to the current timestep (rows 1 and 5), improving the feature extraction ability along the frequency dimension can already improve the transcription accuracy. Our results align with those reported by Kelz \textit{et al.}~\cite{Kelz2016OnTP}, in which they were able to use a relatively simple frame-based model to obtain a relatively good transcription accuracy. Their ConvNet, however, is not entirely frame-based since the convolutional kernel sizes are greater than 1, which allows their model to extract context features. Our experiments extend their results by isolating the temporal features from the model completely, and using the attention to control the exact amount of temporal information accessible to the model.

\subsubsection{Inference model} As mentioned in Section~\ref{sec:related_works}, Onsets and Frames integrates a post-processor to determine the final outcome of the transcription~\cite{Hawthorne2017OnsetsAF}. 
In particular, the model outputs $\hat{\boldsymbol{y}}_{\text{onset}}$ and $\hat{\boldsymbol{y}}_{\text{frame}}$ which are fed to the rule-based inference model $g(\hat{\boldsymbol{y}}_{\text{onset}}\text{, } \hat{\boldsymbol{y}}_{\text{frame}})$ which filters out frames without the onset activation.

This inference model also plays an important role in the achieved transcription accuracy. As shown in Table~\ref{tab:inference}, there is a significant improvement in both note-wise and note-with-offset-wise F1-scores with the inference model. 
It should be pointed out that the rule-based inference model is only beneficial when the accuracy of $\hat{\boldsymbol{y}}_{\text{onset}}$ is reasonably good. 
A noisy $\hat{\boldsymbol{y}}_{\text{onset}}$ would worsen $\hat{\boldsymbol{y}}_{\text{frame}}$. 
In the case of $f_{D=0}$ (without attention, second row) and $f_{D=5}$ (fourth row), the models are too weak to decently predict onsets; we thus use $\hat{\boldsymbol{y}}_{\text{onset}}$ generated by a pre-trained onset stack from Onsets and Frames to demonstrate the effect of the inference model.

One can see that the note-wise and the note-with-offset-wise F1-scores are both improved with the inference model. 
On the other hand, the inference model causes vast degradation in the frame-wise F1-score.
This implies that the inference model acts as a denoising function, removing all the fragmented and redundant notes, causing a large decrease in frame-wise F1-score and a large increase in note-wise F1-score. 
We encourage readers to listen to the transcription results when different model components are missing.\footnote{High resolution figures and transcribed audio samples are available at:\\ \addgitlink}

\subsection{Visualizing Attention Maps}
\label{subsec: attention_maps}
Figure~\ref{fig:attention_map} shows the attention maps for different attended features\footnotemark[3] (each row) for four different input examples. It can be seen from the figure that regardless of which feature is attended to, the attention mechanism always looks for the onset locations (red dotted lines in the figures). Among all features, $\hat{\boldsymbol{y}}_{\text{onset}}$ and $\hat{\boldsymbol{x}}_{\text{spec}}$ yield a much stronger attention than $\hat{\boldsymbol{y}}_{\text{feat}}$. When attending to $\hat{\boldsymbol{y}}_{\text{feat}}$, the attention spreads all over the attention window for most of the time (last row of Figure~\ref{fig:attention_map}). Since $\hat{\boldsymbol{x}}_{\text{spec}}$ is more attentive than $\hat{\boldsymbol{y}}_{\text{feat}}$; and $\hat{\boldsymbol{y}}_{\text{onset}}$ is too sparse and obvious to analyse, we will focus our discussions on $\hat{\boldsymbol{x}}_{\text{spec}}$ in the following paragraphs. We observe the same pattern in other model variations listed in Table~\ref{tab:inference}. To simplify our discussion and save space, we will only discuss the case when the attention mechanism is applied to the complete Onsets and Frames model, but these discussions still hold true in general.

Contrary to the recent belief that including offsets is required for AMT models to perform well~\cite{kim2019adversarial, kelz2019deep}, our results show that the attention mechanism seldom attends to offset locations, unless there is a complete silence after the last note event (yellow dotted line in the 4th column of Figure~\ref{fig:attention_map}). Indeed, introducing an offset sub-module and loss function could result in forcing the model to learn something meaningful and thus boost the transcription performance further. Onset locations seem to be more important than the offset locations as indicated by the attention mechanism.

We can also see that if the music piece has a fast tempo, and the note density is high, the model struggles to find the right place to attend to (2nd column of Figure~\ref{fig:attention_map}). When we decrease the attention window (available in the paper's github page), the attention mechanism starts to pick up the onset location. In this case, the attention mechanism works slightly better with $\hat{\boldsymbol{y}}_{\text{feat}}$, indicating that the convolutional neural network is also extracting useful features for onset locations. Similar findings, that the model is learning beat positions, have been reported before~\cite{ycart2017study, cheukICPR}. The results for other attention maps are available in our demo page.\footnotemark[3]


\section{Conclusion}
\label{sec:conclusion}

In this paper, we revisit the state-of-the-art automatic music transcription model, Onsets and Frames, and try to understand fundamental elements that are essential to produce a high transcription accuracy. Through different experiments conducted in Section~\ref{sec:result}, we discover that (a) various model stacks, (b) moderate amount of temporal information, and (c) the inference model, are the three main components that contrive a good AMT model. Points (a) and (b) are correlated; with a complex enough model, the model can extract suitable amount of temporal information by itself. But adding LSTM layers can explicitly improve point (b). By studying the attention map, we also discover that the onset locations are the most important feature, the final classifier is relying on these to make the prediction as we discussed in Section~\ref{subsec: D_size}. While current research mostly focuses on building very deep and complex models, future research directions should also look into a better way to extract temporal features, or to create better inference models (possibly a neural network-based trainable inference as opposed to rule-based inference).

\section{Acknowledgements}

This work is supported by Singapore International Graduate Award (SINGA) provided by the Agency for Science, Technology and Research (A*STAR) under grant no. SING-2018-02-0204, MOE Tier 2 grant no. MOE2018-T2-2-161, and SRG ISTD 2017 129.

\bibliographystyle{IEEEtran}
\bibliography{refs}

\end{document}

%% file: Tables/compare.tex
\begin{table*}[h]
\centering
 \caption{Results reported as precision (P), recall (R) and F1-score (F1) using the MAPS dataset. To ensure a fair comparison, the Onsets \& Frames model is implemented in PyTorch, which is same as our other models.}
\small
\resizebox{\textwidth}{!}
{\begin{tabular}{l|ccc|ccc|ccc|}
\cline{2-10}
\multicolumn{1}{c|}{}                             & \multicolumn{3}{c|}{\textbf{Frame}} & \multicolumn{3}{c|}{\textbf{Note}}            & \multicolumn{3}{c|}{\textbf{Note w/ offset}}  \\ \cline{2-10} 
\multicolumn{1}{c|}{}                             & P              & R    & F1          & P             & R             & F1            & P             & R             & F1            \\ \hline
\multicolumn{1}{|l|}{Attention on $\hat{\boldsymbol{x}}_{\text{spec}}$}          & 89.4 ± 6.5             & 65.4 ± 9.5 & 75.1 ± 7.2        & 86.3 ± 8.3            & 74.3 ± 11.6          & 79.6 ± 9.7          & 53.2 ± 9.1         &   46.2 ± 11.3        & 49.3 ± 10.2          \\ \hline
\multicolumn{1}{|l|}{Attention on $\hat{\boldsymbol{x}}_{\text{onset}}$} & 89.7 ± 6.2 & 65.7 ± 9.6 & 75.4 ± 7.2 & 85.3 ± 8.5 & 74.6 ± 11.6 & 79.4 ± 9.6 & 53.1 ± 9.5 & 46.8 ± 11.6 & 49.6 ± 10.5 \\ \hline
\multicolumn{1}{|l|}{Attention on $\hat{\boldsymbol{x}}_{\text{feat}}$} & 90.2 ± 5.9 & 64.1 ± 10.1 & 74.5 ± 7.5 & 86.3 ± 8.2 & 73.4 ± 11.5 & 79.0 ± 9.4 & 53.3 ± 9.6 & 45.8 ± 11.8 & 49.1 ± 10.7 \\ \hline
\rowcolor{lightgray} 
\multicolumn{1}{|l|}{\cite{Hawthorne2017OnsetsAF} in \texttt{PyTorch}}     & 90.6 ± 5.8 & 63.1 ± 9.4 & 73.9 ± 7.1 & 85.5 ± 7.7 & 74.1 ± 11.1 & 79.2 ± 9.1 & 52.5 ± 8.9 & 45.8 ± 11.1 & 48.7 ± 10.0  \\ \hline
\end{tabular}}
 \label{tab:compare}
\end{table*}

%% file: Tables/inference.tex
\begin{table}[t]
\centering
 \caption{F1 scores for various metrics on models with and without the rule-based inference.}
\small
\resizebox{\linewidth}{!}
{\begin{tabular}{l|ccc|}
\cline{2-4}
\multicolumn{1}{c|}{}                             & Frame              & Note    & Note w/offset \\ \hline
\multicolumn{1}{|l|}{1. $f_{D=0}$ w/o infer. eq~(\ref{eq:myattention})} & 60.5 ± 11.1& 43.8 ± 11.3& 15.2 ± 7.4     \\ \hline
\multicolumn{1}{|l|}{2. $f_{D=0}$ w/ infer. eq~(\ref{eq:myattention})} & 30.5 ± 12.7& 47.5 ± 16.6& 17.1 ± 8.6     \\ \hline
\multicolumn{1}{|l|}{3. $f_{D=5}$ w/o infer. eq~(\ref{eq:myattention})} & 62.4 ± 12.9 & 47.2 ± 11.2 & 20.4 ± 8.3      \\ \hline
\multicolumn{1}{|l|}{4. $f_{D=5}$ w/ infer. eq~(\ref{eq:myattention})} & 40.9 ± 15.3 & 54.0 ± 17.6 & 23.3 ± 11.5 \\ \hline
\multicolumn{1}{|l|}{5. Conv$_{D=0}$ w/o infer.} & 63.7 ± 9.8& 46.3 ± 10.7& 16.3 ± 7.5 \\ \hline
\multicolumn{1}{|l|}{6. Conv$_{D=0}$ w/ infer.} & 33.2 ± 12.2& 50.4 ± 15.7& 18.4 ± 8.3 \\ \hline
\multicolumn{1}{|l|}{7. Conv$_{D=5}$ w/o infer.} & 66.7 ± 10.6& 49.6 ± 11.2& 20.8 ± 7.9 \\ \hline
\multicolumn{1}{|l|}{8. Conv$_{D=5}$ w/ infer.} & 41.6 ± 14.0& 55.1 ± 15.9& 23.4 ± 10.5 \\ \hline
\multicolumn{1}{|l|}{9. Attn. $\hat{\boldsymbol{x}}_{\text{spec}}$ w/o $F_{\text{onset}}$} & 74.5 ± 6.4& 57.1 ± 11.2& 34.9 ± 10.4       \\ \hline
\multicolumn{1}{|l|}{10. Attn. $\hat{\boldsymbol{x}}_{\text{spec}}$ w/o infer.} & 76.9 ± 6.5 & 65.9 ± 11.0 & 42.4 ± 10.8       \\ \hline
\multicolumn{1}{|l|}{11. Attn. $\hat{\boldsymbol{x}}_{\text{spec}}$ w/o biLSTM} & 65.2 ± 9.5& 75.7 ± 9.7& 40.3 ± 10.6  \\ \hline
\multicolumn{1}{|l|}{12. Attn. $\hat{\boldsymbol{x}}_{\text{spec}}$ w/ infer.} & 75.1 ± 7.2 & 79.6 ± 9.7 & 49.3 ± 10.2  \\ \hline

\rowcolor{lightgray} 
\multicolumn{1}{|l|}{13. \cite{Hawthorne2017OnsetsAF} w/o $F_{\text{onset}}$} & 75.5 ± 6.3& 57.4 ± 11.6& 35.9 ± 10.6      \\ \hline
\rowcolor{lightgray} 
\multicolumn{1}{|l|}{14. \cite{Hawthorne2017OnsetsAF} w/o infer.} & 77.0 ± 6.5 & 65.9 ± 10.9 & 42.4 ± 10.8      \\ \hline
\rowcolor{lightgray} 
\multicolumn{1}{|l|}{15. \cite{Hawthorne2017OnsetsAF} w/o biLSTM} & 63.9 ± 9.2& 74.8 ± 9.5& 39.3 ± 10.6      \\ \hline
\rowcolor{lightgray} 
\multicolumn{1}{|l|}{16. \cite{Hawthorne2017OnsetsAF} w/ infer.} & 73.9 ± 7.1 & 79.2 ± 9.1 & 48.7 ± 10.0 \\ \hline
\end{tabular}}
 \label{tab:inference}
\end{table}

%% file: conference_101719.bbl
\begin{thebibliography}{10}
\providecommand{\url}[1]{#1}
\csname url@samestyle\endcsname
\providecommand{\newblock}{\relax}
\providecommand{\bibinfo}[2]{#2}
\providecommand{\BIBentrySTDinterwordspacing}{\spaceskip=0pt\relax}
\providecommand{\BIBentryALTinterwordstretchfactor}{4}
\providecommand{\BIBentryALTinterwordspacing}{\spaceskip=\fontdimen2\font plus
\BIBentryALTinterwordstretchfactor\fontdimen3\font minus
  \fontdimen4\font\relax}
\providecommand{\BIBforeignlanguage}[2]{{%
\expandafter\ifx\csname l@#1\endcsname\relax
\typeout{** WARNING: IEEEtran.bst: No hyphenation pattern has been}%
\typeout{** loaded for the language `#1'. Using the pattern for}%
\typeout{** the default language instead.}%
\else
\language=\csname l@#1\endcsname
\fi
#2}}
\providecommand{\BIBdecl}{\relax}
\BIBdecl

\bibitem{cuthbert2010music21}
M.~S. Cuthbert and C.~Ariza, ``music21: A toolkit for computer-aided musicology
  and symbolic music data,'' in \emph{ISMIR}, 2010.

\bibitem{huang2020pop}
Y.-S. Huang and Y.-H. Yang, ``Pop music transformer: Generating music with
  rhythm and harmony,'' \emph{arXiv preprint arXiv:2002.00212}, 2020.

\bibitem{herremans2017functional}
D.~Herremans, C.-H. Chuan, and E.~Chew, ``A functional taxonomy of music
  generation systems,'' \emph{ACM Computing Surveys (CSUR)}, vol.~50, no.~5,
  pp. 1--30, 2017.

\bibitem{magalhaeschordify}
J.~P. Magalhaes, ``Chordify: Three years after the launch,'' in \emph{ISMIR},
  2015.

\bibitem{springenberg2014striving}
J.~T. Springenberg, A.~Dosovitskiy, T.~Brox, and M.~Riedmiller, ``Striving for
  simplicity: The all convolutional net,'' \emph{arXiv preprint
  arXiv:1412.6806}, 2014.

\bibitem{Sigtia2015AnEN}
S.~Sigtia, E.~Benetos, and S.~Dixon, ``An end-to-end neural network for
  polyphonic piano music transcription,'' \emph{IEEE/ACM Transactions on Audio,
  Speech, and Language Processing}, vol.~24, pp. 927--939, 2015.

\bibitem{Hawthorne2017OnsetsAF}
C.~Hawthorne, E.~Elsen, J.~Song, A.~Roberts, I.~Simon, C.~Raffel, J.~Engel,
  S.~Oore, and D.~Eck, ``Onsets and frames: Dual-objective piano
  transcription,'' in \emph{ISMIR}, 2017.

\bibitem{kim2019adversarial}
J.~W. Kim and J.~P. Bello, ``Adversarial learning for improved onsets and
  frames music transcription,'' \emph{International Society forMusic
  Information Retrieval Conference}, pp. 670--677, 2019.

\bibitem{kelz2019deep}
R.~Kelz, S.~B{\"o}ck, and G.~Widmer, ``Deep polyphonic adsr piano note
  transcription,'' in \emph{ICASSP 2019-2019 IEEE International Conference on
  Acoustics, Speech and Signal Processing (ICASSP)}.\hskip 1em plus 0.5em minus
  0.4em\relax IEEE, 2019, pp. 246--250.

\bibitem{hawthorne2018enabling}
\BIBentryALTinterwordspacing
C.~Hawthorne, A.~Stasyuk, A.~Roberts, I.~Simon, C.-Z.~A. Huang, S.~Dieleman,
  E.~Elsen, J.~Engel, and D.~Eck, ``Enabling factorized piano music modeling
  and generation with the {MAESTRO} dataset,'' in \emph{International
  Conference on Learning Representations}, 2019. [Online]. Available:
  \url{https://openreview.net/forum?id=r1lYRjC9F7}
\BIBentrySTDinterwordspacing

\bibitem{Kelz18}
R.~Kelz and G.~Widmer, ``Towards interpretable polyphonic transcription with
  invertible neural networks,'' in \emph{ISMIR}, A.~Flexer, G.~Peeters,
  J.~Urbano, and A.~Volk, Eds., 2019, pp. 376--383.

\bibitem{Bahdanau2015NeuralMT}
D.~Bahdanau, K.~Cho, and Y.~Bengio, ``Neural machine translation by jointly
  learning to align and translate,'' \emph{CoRR}, vol. abs/1409.0473, 2015.

\bibitem{luong2015-effective}
T.~Luong, H.~Pham, and C.~D. Manning, ``Effective approaches to attention-based
  neural machine translation,'' in \emph{Proceedings of the 2015 Conference on
  Empirical Methods in Natural Language Processing}.\hskip 1em plus 0.5em minus
  0.4em\relax Lisbon, Portugal: Association for Computational Linguistics, Sep.
  2015, pp. 1412--1421.

\bibitem{XuBKCCSZB15}
K.~Xu, J.~Ba, R.~Kiros, K.~Cho, A.~C. Courville, R.~Salakhutdinov, R.~S. Zemel,
  and Y.~Bengio, ``Show, attend and tell: Neural image caption generation with
  visual attention,'' in \emph{ICML}, 2015, pp. 2048--2057.

\bibitem{emiya2010maps}
V.~Emiya, N.~Bertin, B.~David, and R.~Badeau, ``Maps-a piano database for
  multipitch estimation and automatic transcription of music,'' \emph{Hal
  Inria}, 2010.

\bibitem{cheuk2020impact}
K.~W. Cheuk, K.~Agres, and D.~Herremans, ``The impact of audio input
  representations on neural network based music transcription,''
  \emph{International Joint Conference on Neural Networks}, 2020.

\bibitem{Kelz2016OnTP}
R.~Kelz, M.~Dorfer, F.~Korzeniowski, S.~B{\"o}ck, A.~Arzt, and G.~Widmer, ``On
  the potential of simple framewise approaches to piano transcription,'' in
  \emph{ISMIR}, 2016.

\bibitem{cheuk2019nnaudio}
K.~W. {Cheuk}, H.~{Anderson}, K.~{Agres}, and D.~{Herremans}, ``nnaudio: An
  on-the-fly gpu audio to spectrogram conversion toolbox using 1d convolutional
  neural networks,'' \emph{IEEE Access}, vol.~8, pp. 161\,981--162\,003, 2020.

\bibitem{bay2009evaluation}
M.~Bay, A.~F. Ehmann, and J.~S. Downie, ``Evaluation of multiple-f0 estimation
  and tracking systems.'' in \emph{ISMIR}, 2009, pp. 315--320.

\bibitem{ycart2017study}
A.~Ycart, E.~Benetos \emph{et~al.}, ``A study on lstm networks for polyphonic
  music sequence modelling,'' in \emph{ISMIR}, 2017.

\bibitem{cheukICPR}
K.~W. Cheuk, Y.-J. Luo, E.~Benetos, and D.~Herremans, ``The effect of
  spectrogram reconstructions on automatic music transcription:an alternative
  approach to improve transcription accuracy,'' in \emph{International
  Conference on Pattern Recognition (ICPR 2020)}, in press.

\end{thebibliography}
